\documentclass[12pt]{JHEP3}
\usepackage{axodraw}

\def\Year{\expandafter\eatPrefix\the\year}
\newcount\hours \newcount\minutes
\def\monthname{\ifcase\month\or
January\or February\or March\or April\or May\or June\or July\or
August\or September\or October\or November\or December\fi}
\def\shortmonthname{\ifcase\month\orx
Jan\or Feb\or Mar\or Apr\or May\or Jun\or Jul\or
Aug\or Sep\or Oct\or Nov\or Dec\fi}

\def\TimeStamp{\hours\the\time\divide\hours by60%
\minutes -\the\time\divide\minutes by60\multiply\minutes by60%
\advance\minutes by\the\time
${\rm \shortmonthname}\cdot   \if\day<10{}0\fi\the\day\cdot   \the\year
\qquad\the\hours:\if\minutes<10{}0\fi\the\minutes$}

\newskip\humongous \humongous=0pt plus 1000pt minus 100pt
\def\caja{\mathsurround=0pt}
\def\eqalign#1{\,\vcenter{\openup1\jot \caja
       \ialign{\strut \hfil$\displaystyle{##}$&$
        \displaystyle{{}##}$\hfil\crcr#1\crcr}}\,}
\newif\ifdtup

\newcounter{eqnumber}[section]
\renewcommand{\theeqnumber}{\thesection.\arabic{eqnumber}}
\def\equn{\refstepcounter{eqnumber}
\eqno({\rm \theeqnumber})
}

\def\npb#1#2#3{{\rm Nucl. Phys. B}{\bf \ #1}, #3 (#2)}
\def\plb#1#2#3{{\rm Phys. Lett. B}{\bf \ #1}, #3 (#2)}

\def\hepth#1{[hep-th/#1]}
\def\hepph#1{[hep-ph/#1]}

\newbox\charbox
\newbox\slabox
\def\s#1{{      
        \setbox\charbox=\hbox{$#1$}
        \setbox\slabox=\hbox{$/$}
        \dimen\charbox=\ht\slabox
        \advance\dimen\charbox by -\dp\slabox
        \advance\dimen\charbox by -\ht\charbox
        \advance\dimen\charbox by \dp\charbox
        \divide\dimen\charbox by 2
        \raise-\dimen\charbox\hbox to \wd\charbox{\hss/\hss}
        \llap{$#1$}
}}

\def\spa#1.#2{\left\langle#1\,#2\right\rangle}
\def\spb#1.#2{\left[#1\,#2\right]}
\def\lor#1.#2{\left(#1\,#2\right)}

\catcode`@=11  

\def\Slash#1{\hskip 0.05 cm \slash\hskip -0.22 cm #1}

\def\Tr{\, {\rm Tr}}

\def\la{\langle}
\def\ra{\rangle}

\def\lsl{\not{\hbox{\kern-2.3pt $\ell$}}}
\def\ksl{\not{\hbox{\kern-2.3pt $k$}}}

\def\Z{0}

\def\spa#1.#2{\left\langle#1\,#2\right\rangle}
\def\spb#1.#2{\left[#1\,#2\right]}
\def\lor#1.#2{\left(#1\,#2\right)}

\def\sand#1.#2.#3{%
  \left\langle\smash{#1}{\vphantom1}\right|{#2}%
  \left|\smash{#3}{\vphantom1}\right\rangle}
\def\sandp#1.#2.#3{%
  \left\langle\smash{#1}{\vphantom1}^{-}\right|{#2}%
  \left|\smash{#3}{\vphantom1}^{+}\right\rangle}
\def\sandpp#1.#2.#3{%
  \left\langle\smash{#1}{\vphantom1}^{+}\right|{#2}%
  \left|\smash{#3}{\vphantom1}^{+}\right\rangle}
\def\sandmm#1.#2.#3{%
  \left\langle\smash{#1}{\vphantom1}^{-}\right|{#2}%
  \left|\smash{#3}{\vphantom1}^{-}\right\rangle}
\def\sandpm#1.#2.#3{%
  \left\langle\smash{#1}{\vphantom1}^{+}\right|{#2}%
  \left|\smash{#3}{\vphantom1}^{-}\right\rangle}
\def\sandmp#1.#2.#3{%
  \left\langle\smash{#1}{\vphantom1}^{-}\right|{#2}%
  \left|\smash{#3}{\vphantom1}^{+}\right\rangle}

\def\Atree{A^{\rm tree}}

\def\A#1{{\cal A}_{#1}}

\def\L{\left(}\def\R{\right)}

\def\L{\left(}\def\R{\right)}

\def\tree{{\rm tree}}

\def\Gr{{\rm Gr}}

\def\NeqFour{{\cal N} = 4}
\def\NeqOne{{\cal N} = 1}
\def\NeqZero{{\cal N} = 0}
\def\NeqTwo{{\cal N} = 2}

\def\gluino{\Lambda}
\def\gluinb{\bar\Lambda}

\def\Ang{A_{n}}

\def\npt{$n$-pt }
\def\BRDM#1#2#3#4{\la#1^-|\Slash\hskip -1pt {#2}\Slash\hskip -1pt {#3}|#4^+\ra}
\def\sBRDM#1#2#3#4{\la#1^-|\Slash\hskip 1pt {#2}\Slash\hskip  1pt {#3}|#4^+\ra}

\title{One-Loop NMHV Amplitudes involving Gluinos and Scalars 
       in $\cal{N}$=4 Gauge Theory}
\author{
Steven J. Bidder${}^{1}$,
Warren B. Perkins${}^{1}$ and
Kasper Risager${}^{2}$\\
${}^{1}$Department of Physics,
University of Wales Swansea, Swansea, SA2 8PP, UK\\
${}^{2}$Niels Bohr Institute, University of Copenhagen, 
      Blegdamsvej 17, DK--2100 Copenhagen \o, Denmark }
                                                                               
\preprint{SWAT-05-436}
                                                                               
\abstract{
We use Supersymmetric Ward Identities and quadruple cuts to 
generate $n$-pt NMHV amplitudes involving gluinos and adjoint scalars from
purely gluonic amplitudes. We present a set of factors that can be used 
to generate one-loop NMHV amplitudes involving gluinos or adjoint scalars in
$\NeqFour$ Super Yang-Mills from the corresponding purely gluonic amplitude.
}


\begin{document}

\section{Introduction}

The discovery of a possible duality between gauge theory and twistor
string theory~\cite{Witten:2003nn,CSW}, has led to considerable progress in
obtaining gauge theory amplitudes in compact
forms
~\cite{BDDK7,BDKn,BrittoUnitarity,Cachazo:2004dr,Britto:2004nj,Bidder:2004tx,BBDP2005a,BrittoSQCD}.  While most of
the applications to loop calculations have been to amplitudes which
involve only external gluons, $6$-pt amplitudes involving adjoint fermions and scalars           
have been generated using 
Supersymmetric Ward Identities (SWI)~\cite{Bidder:2005in} and 
superspace constructions~\cite{Huang:2005ve}.

In this paper we use SWI~\cite{SWI} and generalised unitarity cuts~\cite{BrittoUnitarity}
to find one-loop $n$-pt 
``Next-to-Maximally-Helicity-Violating'' or NMHV amplitudes involving  adjoint fermions and scalars
in $\NeqFour$ gauge theory. The amplitudes with purely gluonic external legs were calculated in~\cite{BDKn}
and we present our results as a set of {\it conversion factors} that relate these purely gluonic amplitudes
to those with external fermions/scalars.  
  
In section 2 we review the structure of $\NeqFour$  amplitudes and the recent developments in calculating them in 
compact forms. The purely gluonic amplitudes of ~\cite{BDKn} are reviewed in section 3 and the conversion factors 
calculated in section 4. In section 5 we describe how to compound these factors to find amplitudes with multiple 
external fermions and scalars.  

\section{$\NeqFour$  Amplitudes}

Tree-level amplitudes for $U(N_c)$ or $SU(N_c)$ gauge theories with
$n$ external adjoint particles can be decomposed into colour-ordered partial
amplitudes multiplied by an associated
colour-trace~\cite{TreeColour,ManganoReview}.  Summing over
all non-cyclic permutations reconstructs the full amplitude
$\A{n}^\tree$ from the partial amplitudes $A_n^\tree(\sigma)$,
$$
\A{n}^\tree(\{k_i,a_i\}) =
g^{n-2} \sum_{\sigma\in S_n/Z_n} \Tr(T^{a_{\sigma(1)}}
\cdots T^{a_{\sigma(n)}})
\ A_n^\tree(k_{\sigma(1)}
,\ldots,
            k_{\sigma(n)})
\ ,
\equn\label{TreeAmplitudeDecomposition}
$$
where $k_i$ and $a_i$ are respectively the momentum
and colour-index of the $i$-th external
particle, $g$ is the coupling constant and $S_n/Z_n$ is the set of
non-cyclic permutations of $\{1,\ldots, n\}$.
The $U(N_c)$ ($SU(N_c)$) generators $T^a$ are the set of
traceless hermitian $N_c\times N_c$ matrices,
normalised such that $\Tr\L T^a T^b\R = \delta^{ab}$.
Conventionally we take all particles to be outgoing. We denote gluons 
by $g_i$ and 
adjoint fermions by 
$\Lambda_i$. We will often refer to the adjoint fermions as gluinos for simplicity.

The simplest non-vanishing amplitudes are the 'maximally helicity violating' (MHV)
amplitudes with two particles of negative helicity and the remainder positive. 
The MHV partial amplitudes for gluons are given by the 
Parke-Taylor formulae~\cite{ParkeTaylor}, 
$$
\Atree_n(g_1^+,\ldots,g_j^-,\ldots,g_k^-,
                \ldots,g_n^+)
=\ i\, { {\spa{j}.{k}}^4 \over \spa1.2\spa2.3\cdots\spa{n}.1 }\ ,
\equn\label{ParkeTaylor}
$$ for a partial amplitude where $j$ and $k$ are the legs with
negative helicity.  We use the notation  
$\spa{j}.{l}\equiv \langle j^- | l^+ \rangle $, 
$\spb{j}.{l} \equiv \langle j^+ |l^- \rangle $, 
with $| i^{\pm}\ra $ 
being a massless Weyl spinor with
momentum $k_i$ and chirality
$\pm$~\cite{SpinorHelicity,ManganoReview}.  The spinor products are
related to momentum invariants by 
$\spa{i}.j\spb{j}.i=2k_i \cdot k_j\equiv s_{ij}$ 
with $\spa{i}.j^*=\spb{j}.i$.

The colour decomposition for one-loop amplitudes of adjoint particles
takes the form~\cite{Colour},
$$
{\cal A}_n^{\rm 1-loop}\L \{k_i,a_i\}\R =
g^n \,\sum_{c=1}^{\lfloor{n/2}\rfloor+1}
      \sum_{\sigma \in S_n/S_{n;c}}
     \Gr_{n;c}\L \sigma \R\,A_{n;c}^{}(\sigma),
\label{ColourDecomposition}\equn
$$
where ${\lfloor{x}\rfloor}$ is the largest integer less than or equal to $x$.
The leading colour-structure factor,
$$
\Gr_{n;1}(1) = N_c\ \Tr\L T^{a_1}\cdots T^{a_n}\R \, ,\equn
$$
is just $N_c$ times the tree colour factor, and the subleading colour
structures ($c>1)$ are given by,
$$
\Gr_{n;c}(1) = \Tr\L T^{a_1}\cdots T^{a_{c-1}}\R\,
\Tr\L T^{a_c}\cdots T^{a_n}\R \, .\equn
$$
$S_n$ is the set of all permutations of $n$ objects
and $S_{n;c}$ is the subset leaving $\Gr_{n;c}$ invariant.
Once again it is convenient to use $U(N_c)$ matrices; the extra $U(1)$
decouples~\cite{Colour}.
                                                                               
For one-loop amplitudes the subleading in colour amplitudes
$A_{n;c}$,   $c > 1 $,
may be obtained from summations of permutations
of the leading in colour amplitude~\cite{BDDKa}.
Hence, we need only focus on the leading in colour amplitude $A_{n;1}$, 
which we will generally abbreviate to $A_n$.

One-loop amplitudes depend on the particles circulating in the
loop.  In $\NeqFour$ SYM cancellations between fermionic loops and bosonic loops
lead to considerable simplifications in the loop momentum integrals. This is
manifest in the ``string-based approach'' to computing loop
amplitudes~\cite{StringBased}.  As a result,
$\NeqFour$ one-loop amplitudes can be expressed simply as a sum of
scalar box-integral functions~\cite{BDDKa},
 $$
  I_{i}^{1m}
\hskip 0.5truecm
 I_{r;i}^{2me}
\hskip 0.5truecm
  I_{r;i}^{2mh}
\hskip 0.5truecm
  I_{r,r',i}^{3m}
\hskip 0.5truecm
I_{r, r', r'', i}^{4m} \equn
$$
with the labeling as indicated,
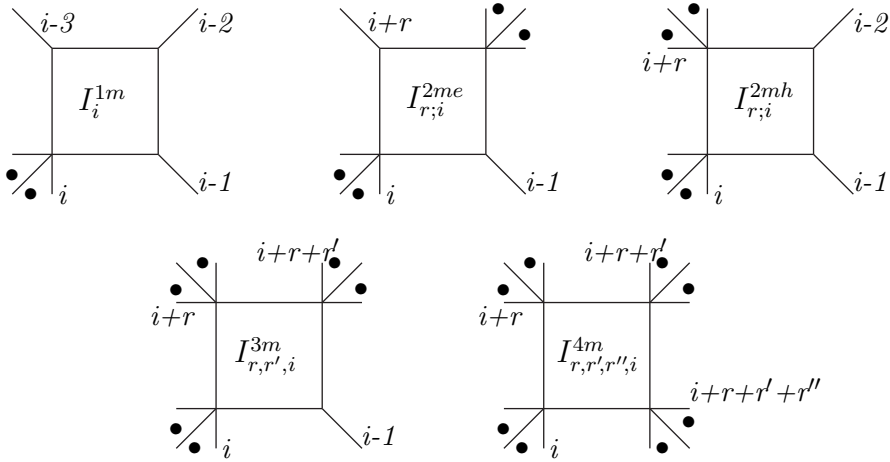
\begin{center}
\begin{picture}(120,95)(0,0)
\Line(30,20)(70,20)
\Line(30,60)(70,60)
\Line(30,20)(30,60)
\Line(70,60)(70,20)
                                                                                
\Line(30,20)(15,5)
\Line(30,60)(15,75)
\Line(70,20)(85,5)
\Line(70,60)(85,75)
                                                                                
\Line(30,20)(30,5)
\Line(30,20)(15,20)
\Text(22,5)[c]{\small$\bullet $}
\Text(15,12)[c]{\small $\bullet $}
                                                                                
\Text(32,5)[l]{\small $\hbox{\it i}$}
\Text(85,10)[l]{\small $\hbox{\it i-1}$}
\Text(85,70)[l]{\small $\hbox{\it i-2}$}
\Text(25,70)[l]{\small $\hbox{\it i-3}$}

\Text(40,40)[l]{$I^{1m}_{i}$}
\end{picture}
\begin{picture}(120,95)(0,0)
\Line(30,20)(70,20)
\Line(30,60)(70,60)
\Line(30,20)(30,60)
\Line(70,60)(70,20)
                                                                                
\Line(30,20)(15,5)
\Line(30,60)(15,75)
\Line(70,20)(85,5)
\Line(70,60)(85,75)
                                                                                
\Line(30,20)(30,5)
\Line(30,20)(15,20)
\Text(22,5)[c]{\small$\bullet$}
\Text(15,12)[c]{\small $\bullet$}
                                                                                
\Line(70,60)(70,75)
\Line(70,60)(85,60)
\Text(75,75)[c]{\small$\bullet$}
\Text(85,65)[c]{\small $\bullet$}
\Text(32,5)[l]{\small $\hbox{\it i}$}
\Text(85,10)[l]{\small $\hbox{\it i-1}$}
\Text(25,70)[l]{\small $\hbox{\it i+r}$}

\Text(40,40)[l]{$I^{2me}_{r;i}$}
\end{picture}
\begin{picture}(120,95)(0,0)
\Line(30,20)(70,20)
\Line(30,60)(70,60)
\Line(30,20)(30,60)
\Line(70,60)(70,20)
                                                                                
\Line(30,20)(15,5)
\Line(30,60)(15,75)
\Line(70,20)(85,5)
\Line(70,60)(85,75)
                                                                                
\Line(30,20)(30,5)
\Line(30,20)(15,20)
\Text(22,5)[c]{\small$\bullet$}
\Text(15,12)[c]{\small $\bullet$}
                                                                                
\Line(30,60)(30,75)
\Line(30,60)(15,60)
\Text(15,65)[c]{\small$\bullet$}
\Text(25,75)[c]{\small $\bullet$}

\Text(32,5)[l]{\small $\hbox{\it i}$}
\Text(85,10)[l]{\small $\hbox{\it i-1}$}
\Text(85,70)[l]{\small $\hbox{\it i-2}$}
\Text(5,55)[l]{\small $\hbox{\it i+r}$}

\Text(40,40)[l]{$I^{2mh}_{r;i}$}
\end{picture}
                                                                           
\begin{picture}(120,95)(0,0)
\Line(30,20)(70,20)
\Line(30,60)(70,60)
\Line(30,20)(30,60)
\Line(70,60)(70,20)
                                                                                
\Line(30,20)(15,5)
\Line(30,60)(15,75)
\Line(70,20)(85,5)
\Line(70,60)(85,75)
                                                                                
\Line(30,20)(30,5)
\Line(30,20)(15,20)
\Text(22,5)[c]{\small $\bullet$}
\Text(15,12)[c]{\small $\bullet$}
                                                                                
\Line(30,60)(30,75)
\Line(30,60)(15,60)
\Text(15,65)[c]{\small $\bullet$}
\Text(25,75)[c]{\small $\bullet$}

\Line(70,60)(70,75)
\Line(70,60)(85,60)
\Text(75,75)[c]{\small $\bullet$}
\Text(85,65)[c]{\small $\bullet$}
                                                                                                                                                                
\Text(32,5)[l]{\small $\hbox{\it i}$}
\Text(85,10)[l]{\small $\hbox{\it i-1}$}
\Text(45,80)[l]{\small $\hbox{\it i+r+r}'$}
\Text(5,55)[l]{\small $\hbox{\it i+r}$}

\Text(37,40)[l]{$I^{3m}_{r,r',i}$}
\end{picture}
\begin{picture}(120,95)(0,0)
\Line(30,20)(70,20)
\Line(30,60)(70,60)
\Line(30,20)(30,60)
\Line(70,60)(70,20)
                                                                                
\Line(30,20)(15,5)
\Line(30,60)(15,75)
\Line(70,20)(85,5)
\Line(70,60)(85,75)
                                                                                
\Line(30,20)(30,5)
\Line(30,20)(15,20)
\Text(22,5)[c]{\small $\bullet$}
\Text(15,12)[c]{\small $\bullet$}
                                                                                
\Line(30,60)(30,75)
\Line(30,60)(15,60)
\Text(15,65)[c]{\small $\bullet$}
\Text(25,75)[c]{\small $\bullet$}
                                                                                                                                                                
\Line(70,60)(70,75)
\Line(70,60)(85,60)
\Text(75,75)[c]{\small $\bullet$}
\Text(85,65)[c]{\small $\bullet$}
                                                                                
\Line(70,20)(70,5)
\Line(70,20)(85,20)
\Text(75,5)[c]{\small $\bullet$}
\Text(85,15)[c]{\small $\bullet$}

\Text(32,5)[l]{\small $\hbox{\it i}$}
\Text(85,27)[l]{\small $\hbox{\it i+r+r}'\hbox{\it +r}''$}
\Text(45,80)[l]{\small $\hbox{\it i+r+r}'$}
\Text(5,55)[l]{\small $\hbox{\it i+r}$}

\Text(35,40)[l]{$I^{4m}_{r,r'\hskip -2pt ,r''\hskip -2pt,i}$}
\end{picture}

\end{center}
                                                                                
Explicit forms for these scalar box integrals can be found in ~\cite{BDDKa}. 
It is convenient to
define dimension zero $F$-functions, $F_i$, by removing the momentum
prefactors of these scalar boxes~\cite{BDDKb}.
The one-loop amplitudes can then  be  expressed as, 
$$
A^{\NeqFour}  = \sum_{i}  c_i F_i \, ,
\equn\label{generalform2}
$$
and the computation of one-loop $\NeqFour$ amplitudes is just a matter of 
determining the coefficients $c_i$.
These remarkable simplifications also appear to extend beyond one-loop~\cite{MultiLoop}. 

These $\NeqFour$ amplitudes are related to purely gluonic amplitudes via a sum
over contributions from various matter supermultiplets:
$$
A_{n}\ \equiv\
A_{n}^{\;\NeqFour}-4A_{n}^{\;\NeqOne\ {\rm chiral}} 
 + A_{n}^{[0]}\, ,
\equn
$$
where $A_{n}^{[0]}$ is the contribution from the complex scalar (or $\NeqZero$ matter multiplet) 
circulating in the loop. 
(Throughout we assume the use of a supersymmetry preserving
regulator~\cite{Siegel,StringBased,KST}.)

Progress in calculating amplitudes has been remarkable and varied.  At
tree level, inspired by the duality between topological string theory
and gauge theory~\cite{Witten:2003nn}, a reformulation of perturbation
theory in terms of MHV-vertices was proposed~\cite{CSW}. This promoted
the MHV amplitudes of eq.~(\ref{ParkeTaylor}) to the role of fundamental
building blocks in the perturbative expansion. By continuing legs
off-shell in a well specified manner these can be sewn together to
form other amplitudes.  This reformulation
produces relatively compact expressions for tree
amplitudes. While initially presented for purely gluonic amplitudes,
it has been successfully extended to other particle types
~\cite{CSW:matter,CSW:massive}.

In a different development,
a series of recursion relations for calculating tree
amplitudes have been introduced~\cite{Britto:2004ap}. These yield
compact expressions for gluonic tree amplitudes~\cite{Britto:2005dg}, the
six-point NMHV amplitudes involving fermions~\cite{Luo:2005rx}
and gravity amplitudes~\cite{Bedford:2005yy,Cachazo:2005ca,Bjerrum-Bohr:2005xx}.

For one-loop amplitudes in gauge theory, full results for all
helicities and all particle types are only known for the
four-point~\cite{EllisSexton,KST} and five-point~\cite{FiveGluon,Bern:1994fz,Kunszt:1994tq,Glover:1996eh}
amplitudes. Recently recursive techniques have been used to obtain certain $6$-pt and $7$-pt MHV
one-loop amplitudes in QCD~\cite{Bern:2005cq}. 
  Beyond five-point, the one-loop amplitudes are much
better understood within supersymmetric theories.  
Here the amplitudes are ``cut-constructible'',
in that the coefficients can be determined from unitarity cuts. 
Using this fact, in ~\cite{BDDKa} the one-loop amplitudes 
were determined for the all-$n$
MHV amplitudes and in ~\cite{BDDKb} the remaining six-point gluonic amplitudes
(the NMHV amplitudes) were computed and the MHV amplitudes determined in 
$\NeqOne$ theories. Recursive techniques have been used to generate the $n$-pt one-loop
split-helicity amplitudes~\cite{Bern:2005hh}.

The MHV vertex approach has also been
shown to extend to one-loop in ~\cite{Brandhuber:2004yw}, where the
one-loop $\NeqFour$ MHV amplitudes were computed and shown to be in complete
agreement with the results of ~\cite{BDDKa}, and in
~\cite{Quigley:2004pw,Bedford:2004py} where the $\NeqOne$ MHV
one-loop amplitudes were computed and shown to be in agreement with the results
of~\cite{BDDKb}.  

These techniques have been very successful and  results
include the recent computation of all $\NeqFour$ NMHV
one-loop amplitudes with external glue~\cite{BDDK7,BDKn} and 
various next-to-next-to-MHV (N$^2$MHV)
box-coefficients~\cite{BrittoUnitarity}. 

An important development which enhances the power of the unitarity method, is
the observation by Britto, Cachazo and Feng~\cite{BrittoUnitarity} that
box integral coefficients can be obtained from generalised unitarity
cuts~\cite{Eden,GeneralizedCuts,BDDK7} by analytically 
continuing the massless corners of 
the quadruple cuts.  
The quadruple cuts give each
box-coefficient as a product of four tree amplitudes.
Applying this to the box,
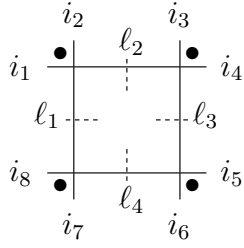
\begin{center}
\begin{picture}(160,100)(0,0)
\DashLine(50,73)(50,61){2}
\DashLine(50,39)(50,27){2}
\DashLine(27,50)(39,50){2}
\DashLine(61,50)(73,50){2}

\Line(30,30)(30,70)
\Line(70,30)(70,70)
\Line(30,30)(70,30)
\Line(70,70)(30,70)

\Line(30,70)(20,70)
\Line(30,70)(30,80)

\Line(70,30)(70,20)
\Line(70,30)(80,30)

\Line(70,70)(70,80)
\Line(70,70)(80,70)

\Line(30,20)(30,30)
\Line(20,30)(30,30)

\Text(30,10)[c]{$i_7$}
\Text(10,30)[c]{$i_8$}
\Text(25,25)[c]{$\bullet$}
\Text(70,10)[c]{$i_6$}
\Text(90,30)[c]{$i_5$}\Text(70,90)[c]{$i_3$}
\Text(90,70)[c]{$i_4$}
\Text(30,90)[c]{$i_2$}
\Text(10,70)[c]{$i_1$}
\Text(75,25)[c]{$\bullet$}
\Text(75,75)[c]{$\bullet$}
\Text(25,75)[c]{$\bullet$}

\Text(20,52)[c]{$\ell_1$}
\Text(52,20)[c]{$\ell_4$}
\Text(80,52)[c]{$\ell_3$}
\Text(52,80)[c]{$\ell_2$}
\end{picture}
\label{QuadrupleCutFigure}
\end{center}
where dashed lines represent cuts and dots represent arbitrary numbers of
external line insertions, the box-coefficient is given by, 
$$
\eqalign{
c={ 1 \over 2 } \sum_{\cal S}
\biggl( \Atree(\ell_1,i_1,  & \ldots,i_2,\ell_2) \times
\Atree(\ell_2,i_3,\ldots,i_4,\ell_3)
\cr
& \times \Atree(\ell_3,i_5,\ldots,i_6,\ell_4) \times
\Atree(\ell_4,i_7,\ldots,i_8,\ell_1) \biggr)\;,
\cr}
\equn\label{QuadCuts}
$$ 
where the sum is over all allowed intermediate configurations and
particle types~\cite{BrittoUnitarity} and the cut legs are frozen
in a specific manner. 

These techniques are also useful in calculating amplitudes in 
${\cal N} < 4$ theories \cite{Bidder:2004tx,BBDP, BBDP2005a,BrittoSQCD},  
although
these amplitudes are more complicated and contain
integral functions other than the box functions. 
Unfortunately, non-supersymmetric theories are not cut-constructible~\cite{BDDKb}, so the unitary
techniques are not immediately applicable, although progress is ongoing in 
this area~\cite{Bern:2005hs,Bern:2005ji}. 

Amplitudes with external glue can be related to ones with external fermions and scalars
using Supersymmetric Ward Identities (SWI)~\cite{Bidder:2005in}. The SWI
 can be obtained by acting with the
supersymmetry generator $Q(\eta)$ (where $\eta$ is an arbitrary spinor parameter)
on a string of operators, $z_i$, which has
vanishing vacuum expectation value.  
Since $Q(\eta)$ annihilates the vacuum we obtain,
$$
0=  \biggl< \Bigl[ Q(\eta) , \prod_{i} z_i \Bigr] \biggr>=
\sum_i \Bigl< z_1\cdots [Q(\eta),z_i]\cdots z_n \Bigr>\, . 
\equn
$$
\noindent
For $\NeqOne$ supersymmetry we can use the supersymmetry algebra, 
$$
\eqalign{
[Q(\eta) , g^{+} (p)] =- \spb{\eta}.{p} \gluinb^{+},\quad &
[Q(\eta) , g^{-} (p)] =  \spa{p}.{\eta} \gluino^{-},
\cr
[Q(\eta) , \gluinb^{+} (p)] =- \spa{p}.{\eta} g^{+},\quad &
[Q(\eta) , \gluino^{-} (p)] =  \spb{\eta}.{p} g^{-},
\cr}
\equn$$
where $g^{\pm}(p)$ is the operator creating a gluon of momentum $p$ and
$\gluino^{-}(p)$ that for a gluino.
Applying this to $\Ang'( g_1^-,g_2^- , \gluino_3^+,g_4^+,\ldots, g_n^+)$ 
we obtain,
$$
\eqalign{
0=\spa1.{\eta} \Ang( \gluino_1^-,g_2^-,  \gluinb_3^+,g_4^+,\ldots, g_n^+)
+\spa2.{\eta} \Ang( g_1^-,\gluino_2^-, \gluinb_3^+,g_4^+,\ldots, g_n^+)
\cr
-\spa3.{\eta} \Ang( g_1^-,g_2^-,  g_3^+,g_4^+,\ldots, g_n^+)
\cr},
\equn$$ 
where we have used the fact that amplitudes with two fermions of the 
same helicity vanish. Appropriate choices of $\eta$ then give the MHV 
two-gluino amplitudes directly in terms of the purely gluonic amplitude.  

For NMHV amplitudes there are typically four amplitudes contributing to each identity.
The system has rank 2, so it can only directly give two of the
amplitudes in terms of the other two. In \cite{Bidder:2005in} symmetry arguments were
used to resolve these ambiguities and solve the SWI at $6$-pt. Alternatively, a second amplitude
can be calculated explicitly and the SWI used to generate the other two. This is the approach we 
adopt in this paper. SWI apply order by order in perturbation theory and, as the box-integral 
functions are a set of independent functions, box by box.    

Recently very elegant expressions for
the $6$-pt amplitudes have been derived using a superspace construction~\cite{Huang:2005ve}.

\section{Summary of NMHV Gluonic Amplitudes}

In this section we review the \npt  one-loop 
NMHV gluonic amplitudes derived in 
\cite{BDKn}. Our gluino amplitudes will
be derived from these.

In any one-loop NMHV box diagram there are seven legs with 
negative helicity: three external and four internal. As each
massive corner of the box requires at least two legs with 
negative helicity to be non-zero, we can have at most three 
massive corners. Further, the three-mass boxes have a particularly
simple form, with three massive MHV corners and one massless $\overline{\rm MHV}$ (or googly)
corner. Thus they are 'MHV-deconstructible' in that they can be determined 
using purely MHV vertices and, in this case, quadruple cuts. The three-mass box-coefficient
$c^{3m}(m_1,m_2,m_3;A,B,C,d)$ where $A$, $B$ and $C$ are the massive corners,
$d$ is the massless corner and $m_1$, $m_2$ and $m_3$ are the legs with 
negative helicity, is given by~\cite{BDKn},
$$
\array{lc}
c^{3m}(m_1,m_2,m_3;A,B,C,d)= & \\
\,\,\,\,\,\,\,\,
{
[{\cal H}(m_1,m_2,m_3;A,B,C,d)]^4
\over 
\spa{1}.{2}\spa{2}.{3}\ldots \spa{n}.{1}K^2_B
}
{
\spa{A_{-1}}.{B_1}
\over
\BRDM{d}{K_C}{K_B}{A_{-1}}\BRDM{d}{K_C}{K_B}{B_1}
}
{
\spa{B_{-1}}.{C_1}
\over
\BRDM{d}{K_A}{K_B}{B_{-1}}\BRDM{d}{K_A}{K_B}{C_1}
},
&\\
\endarray
\equn
$$
where $A_1$ denotes the first leg of corner $A$ and $A_{-1}$ the last. 
When leg $d$ has positive helicity, ${\cal H}$ is given by,
$$
\array{lll}
{\cal H}  & =0,                                 &  {m_{1,2,3}\in A,}             \\
          & =0,                                 &  {m_{1,2,3}\in B,}             \\
          & =\spa{m_1}.{m_2}\BRDM{d}{K_C}{K_B}{m_3},   & {m_{1,2}\in A, \,  m_{3}\in B,}\\
          & =\spa{m_2}.{m_3}\BRDM{d}{K_C}{K_B}{m_1},   & {m_{1}\in A, \,  m_{2,3}\in B,}\\
          & =\spa{m_1}.{m_2}\spa{d}.{m_3}K_B^2,              & {m_{1,2}\in A, \,  m_{3}\in C,}\\
          & =\spa{m_1}.{m_2}\BRDM{d}{K_A}{K_B}{m_3}   &                                 \\
          & \, \, \, \,\, +\spa{m_3}.{m_2}\BRDM{d}{K_C}{K_B}{m_1}, \, \, \, \,\,   & {m_{1} \in A, \, m_2\in B, \, m_{3}\in C,}\\
\endarray
\equn$$
and when leg $d$ has negative helicity, $d=m_3$, ${\cal H}$ is given by,
$$
\array{lll}
{\cal H}  & =0,                                 &  {m_{1,2}\in A,}             \\
          & =\spa{m_1}.{m_2}\BRDM{d}{K_C}{K_B}{d},     & {m_{1,2}\in B, }              \\
          & =\spa{d}.{m_1}\BRDM{d}{K_C}{K_B}{m_2},   & {m_{1}\in A, \,  m_{2}\in B,}\\
          & =\spa{d}.{m_1}\spa{d}.{m_2}K_B^2,              & {m_{1}\in A, \,  m_{2}\in C.}\\
\endarray
\equn$$

The two-mass hard boxes are also MHV-deconstructible. As boxes with adjacent massless corners of 
the same type vanish, each non-vanishing two-mass easy box has a massless MHV corner, 
a massless googly corner and two massive MHV corners. Unfortunately, two-mass easy and one-mass
boxes are not all MHV-deconstructible. However, all three types of box can be generated from
the three-mass boxes using IR consistency arguments~\cite{BDKn}. For the two-mass hard boxes
the result is,
$$
c^{2mh}(A,B,c,d)=c^{3m}(A,B,\{c\},d)+c^{3m}(A,B,c,\{d\}),
\equn$$
where lower case letters denote massless corners and $\{\}$ indicates that the corner should be
thought of as the massless limit of a massive corner. This relationship has a simple interpretation
in terms of the box diagrams: for each internal helicity configuration, one of the massless corners
of the two-mass hard box will be MHV and can be thought of as the massless
limit of a massive MHV corner. Summing over internal helicity configurations in general gives two 
terms. If one of the helicity configurations gives a vanishing contribution, the corresponding
three-mass box-coefficient will also vanish.    

The two-mass easy boxes can also be expressed in terms of three-mass boxes~\cite{BDKn}:
$$
c^{2me}(A,b,C,d)=\sum_{\hat X,\hat Y,\hat Z} c^{3m}(b,\hat X(d),\hat Y,\hat Z)
                    +\sum_{\hat X,\hat Y,\hat Z} c^{3m}(d,\hat X(b),\hat Y,\hat Z),
\equn$$
where the sum is over all clusters (maintaining cyclic ordering) where $\hat X(a)$
contains leg $a$ and $\hat Y$ is massive. Finally, the one-mass boxes are given by,
$$
c^{1m}(A,b,c,d)=c^{2me}(A,b,\{c\},d)+c^{3m}(A,\{b\},c,\{d\}).
\equn$$
These relationships are based on the IR properties of the box integral functions and thus carry over
directly to amplitudes involving gluinos and scalars.

\section{Conversion Factors From Supersymmetric Ward Identities and Quadruple Cuts}   

We first consider amplitudes with a pair of external gluinos. These are related to
purely gluonic amplitudes through the SWI obtained by acting with 
supersymmetry generator $Q$ on $A(g_{m_1}^-,g_{m_2}^-,g_{m_3}^-,\gluinb_q^+,\ldots)$, where 
$\ldots$ represents a string of positive helicity gluons.
The structure of the SWI is independent of the ordering of the legs, but there are different
SWI for each distinct ordering. We will explicitly show the case where the first three legs 
have negative helicity.  As the SWI apply box by box, we have, 
$$
\eqalign{
\la q\eta\ra c(g_{m_1}^-,g_{m_2}^-,g_{m_3}^-,g_q^+,\ldots)=
&\,\,\,\,\,\,\la m_1\eta\ra c(\gluino_{m_1}^-,g_{m_2}^-,g_{m_3}^-,\gluinb_q^+,\ldots)\cr
&+\la m_2\eta\ra c(g_{m_1}^-,\gluino_{m_2}^-,g_{m_3}^-,\gluinb_q^+,\ldots)\cr
&+\la m_3\eta\ra c(g_{m_1}^-,g_{m_2}^-,\gluino_{m_3}^-,\gluinb_q^+,\ldots),
}\equn\label{eq:nmhvswi0}
$$
where $c$ is a generic box-coefficient.

The SWI (\ref{eq:nmhvswi0}) has rank two, so it determines two of the box-coefficients  in terms of the
other two. Our approach is to determine one of the two-gluino box-coefficients  using quadruple cuts and
then use the SWI to generate the other two. As in the purely gluonic case, we can express 
all of our box-coefficients  as sums of the three-mass ones, so we only need to evaluate the latter explicitly.
As the three-mass boxes are MHV-deconstructible, we can determine any tree amplitude we need using 
\cite{Nair}. A generic box is shown in the figure: 

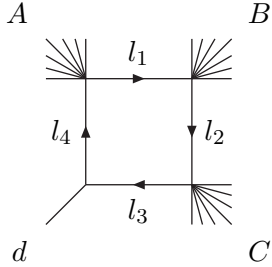
\begin{center}
\begin{picture}(120,100)(0,-5)
%
\ArrowLine(30,60)(70,60)
\Text(50,70)[c]{\small $l_1$}
%
\ArrowLine(70,60)(70,20)
\Text(75,40)[l]{\small $l_2$}
%
\ArrowLine(70,20)(30,20)
\Text(50,10)[c]{\small $l_3$}
%
\ArrowLine(30,20)(30,60)
\Text(25,40)[r]{\small $l_4$}
%
\Text(10,85)[c]{{\small $A$}\quad}
\Line(30,60)(15,60)
\Line(30,60)(15,64)
\Line(30,60)(15,68.7)
\Line(30,60)(15,75)
\Line(30,60)(21.3,75)
\Line(30,60)(26,75)
\Line(30,60)(30,75)
%
\Text(90,85)[c]{\quad\small $B$}
\Line(70,60)(70,75)
\Line(70,60)(74,75)
\Line(70,60)(78.7,75)
\Line(70,60)(85,75)
\Line(70,60)(85,68.7)
\Line(70,60)(85,64)
\Line(70,60)(85,60)
%
\Text(90,-5)[c]{\quad\small $C$}
\Line(70,20)(85,20)
\Line(70,20)(85,16)
\Line(70,20)(85,11.3)
\Line(70,20)(85,5)
\Line(70,20)(78.7,5)
\Line(70,20)(74,5)
\Line(70,20)(70,5)
%
\Text(5,-5)[c]{\small $d$}
\Line(30,20)(15,5)
\end{picture}
\end{center}
The massless corner is a 3-pt googly vertex, so we have the following useful results,
$$
l_3=l_4+d,
\qquad 
|l^+_3\ra \propto |d^+\ra,
\qquad 
|l^+_4\ra \propto |d^+\ra,
$$
$$
\la dl_1\ra[l_1l_2]\la l_2|=\la d^-|\Slash\hskip -2pt K_A\Slash\hskip -2pt K_B|,
\qquad 
\la dl_2\ra[l_2l_1]\la l_1|=\la d^-|\Slash\hskip -2pt K_C\Slash\hskip -2pt K_B|,
\equn$$
which allow us to evaluate expressions that are homogeneous in $l_i$.

We label the purely gluonic boxes by the location of the negative helicity legs. The 'AAB' box shown 
below has two negative helicity legs on corner A and one on corner B:
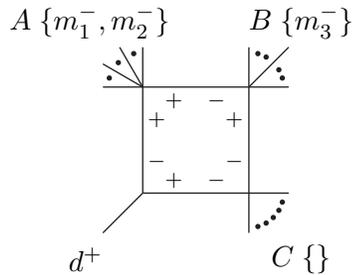
\begin{center}
\begin{picture}(120,100)(0,-5)
%
\Line(30,60)(70,60)
\Text(42,55)[c]{\scriptsize $+$}\Text(58,55)[c]{\scriptsize $-$}
%
\Line(70,60)(70,20)
\Text(68,48)[r]{\scriptsize $+$}\Text(68,32)[r]{\scriptsize $-$}
%
\Line(70,20)(30,20)
\Text(58,25)[c]{\scriptsize $-$}\Text(42,25)[c]{\scriptsize $+$}
%
\Line(30,20)(30,60)
\Text(32,32)[l]{\scriptsize $-$}\Text(32,48)[l]{\scriptsize $+$}
%
\Text(10,85)[c]{\small $A \; \{m_1^-,m_2^-\}$}
\Line(30,60)(15,60)
\Line(30,60)(15,68.7)
\Line(30,60)(21.3,75)
\Line(30,60)(30,75)
\Vertex(17.4,63.4){1}
\Vertex(20.8,69.2){1}
\Vertex(26.6,72.6){1}
%
\Text(90,85)[c]{\small $B \; \{m_3^-\}$}
\Line(70,60)(70,75)
\Line(70,60)(85,75)
\Line(70,60)(85,60)
\Vertex(73.4,72.6){1}
\Vertex(76.5,71.3){1}
\Vertex(81.3,66.5){1}
\Vertex(82.6,63.4){1}
%
\Text(90,-5)[c]{\small $C \; \{\}$}
\Line(70,20)(85,20)
\Line(70,20)(70,5)
\Vertex(82.6,16.6){1}
\Vertex(81.3,13.5){1}
\Vertex(79.2,10.8){1}
\Vertex(76.5,8.7){1}
\Vertex(73.4,7.4){1}
%
\Text(10,-5)[c]{\small $d^+ \; $}
\Line(30,20)(15,5)
\end{picture}
\end{center}
When we relate this purely gluonic box to one with a pair of external gluinos, we must 
specify which $g^+$ to replace by $\gluinb^+$ and which  $g^-$ to replace by $\gluino^-$.
Using  $m$ to label the external $\gluino^-$ leg,
       $q$ to label the external $\gluinb^+$ leg and  $L(q)$ to denote its location, 
the conversion factor is given by $R^{\rm box \; label}_{L(q)m}$, so that,
$$
c^{xxx}(g^-,g^-,\gluino_{m}^-,\gluinb_q^+,\ldots)=R^{xxx}_{L(q)m}c^{xxx}(g^-,g^-,g_{m}^-,g_q^+,\ldots).
\equn$$

The AAB box shown is an example of a 'singlet' box, where only gluons can circulate in the loop and
there is a single contribution to the purely gluonic box-coefficient.  We can immediately see that there is
no possible routing for a fermion, $\gluino^-$, from corner B to corner A or d, so we have,
$$
R^{AAB}_{Am_3}=R^{AAB}_{dm_3}=0.
\equn$$
The remaining $R^{AAB}_{Am_i}$ and $R^{AAB}_{dm_i}$ then follow from the SWI. For $R^{AAB}_{Bm_i}$ and $R^{AAB}_{Cm_i}$
there are either one or two possible fermion routings and one box in each class must be calculated using quadruple cuts before
the other two can be read off from the SWI. 
The conversion factors for the other singlet boxes can be similarly evaluated.  
The results of these calculations are presented in table 1.

The ABC boxes are 'non-singlet' and any particle in the $\NeqFour$ multiplet can circulate in the loop.
The purely gluonic box-coefficients  are obtained by summing over diagrams with all possible particles circulating in the loop.
If the  $\gluinb^+$ and $\gluino^-$ attach to the same corner, any particle can still circulate in the loop. Each
corner remains MHV, but care must be taken with the flavour structure of corners with four non-gluonic legs as these amplitudes
are flavour dependent. In all cases the MHV tree amplitudes can be found using ~\cite{Nair}.
 
Our results are presented in table 1. For each type of box the conversion factors have a common denominator. The factor
appearing in each denominator also appears in the numerator of the corresponding purely gluonic amplitude, where it is raised to the
fourth power. Conversion factors are presented for all distinct cases. The factors not explicitly listed can be obtained by
flipping (e.g. AAB boxes flip into BCC boxes). The denominator of each conversion factor is given next to the box name and
the numerators are listed for each location of $q$ and for each $m$.

\begin{center}
\scriptsize
%
%
\begin{tabular}{|c|c|r|}
\hline
\hline
\multicolumn{2}{|c|}{AAB}&$\sBRDM d{K_C}{K_B}{m_3}\la m_1m_2\ra$\\
\hline
\hline
$A$&$m_1$&$\sBRDM d{K_C}{K_B}{m_3}\la q m_2\ra$\\
$/d$&$m_2$&$\sBRDM d{K_C}{K_B}{m_3}\la m_1 q\ra$\\
&$m_3$&$0$\\
\hline
$B$&$m_1$&$-\sBRDM d{K_C}{K_B}{m_2}\la m_3q\ra$\\
&$m_2$&$-\sBRDM d{K_C}{K_B}{m_1}\la qm_3\ra$\\
&$m_3$&$\sBRDM d{K_C}{K_B}{q}\la m_1m_2\ra$\\
\hline
$C$&$m_1$&$\sBRDM d{K_A}{K_B}{q}\la m_2m_3\ra-\sBRDM d{K_C}{K_B}{m_2}\la m_3q\ra$\\
&$m_2$&$\sBRDM d{K_A}{K_B}{q}\la m_3m_1\ra-\sBRDM d{K_C}{K_B}{m_1}\la qm_3\ra$\\
&$m_3$&$\sBRDM d{K_A}{K_B}{q}\la m_1m_2\ra+\sBRDM d{K_C}{K_B}{q}\la m_1m_2\ra$\\
&&$=-K_B^2\la dq\ra\la m_1m_2\ra$\\
\hline
\hline
\end{tabular}~~
%
%

\end{center}
\begin{center}\scriptsize
%
%
\begin{tabular}{|c|c|r|}
\hline
\hline
\multicolumn{2}{|c|}{AAC}&$K_B^2\la dm_3\ra\la m_1m_2\ra$\\
\hline
\hline
$A$&$m_1$&$K_B^2\la dm_3\ra\la q m_2\ra$\\
$/d$&$m_2$&$K_B^2\la dm_3\ra\la m_1 q\ra$\\
&$m_3$&$0$\\
\hline
$B$&$m_1$&$-\sBRDM d{K_A}{K_B}{m_3}\la qm_2\ra
+\sBRDM d{K_C}{K_B}{m_2}\la m_3q\ra$\\
&$m_2$&$-\sBRDM d{K_A}{K_B}{m_3}\la m_1q\ra
+\sBRDM d{K_C}{K_B}{m_1}\la qm_3\ra$\\
&$m_3$&$-\sBRDM d{K_C}{K_B}{q}\la m_1m_2\ra$\\
\hline
$C$&$m_1$&$-K_B^2\la dm_2\ra\la m_3q\ra$\\
&$m_2$&$-K_B^2\la dm_1\ra\la qm_3\ra$\\
&$m_3$&$K_B^2\la dq\ra\la m_1m_2\ra$\\
\hline
\hline
\end{tabular}~~
%
%

\end{center}
\begin{center}\scriptsize
%
%
\begin{tabular}{|c|c|r|}
\hline
\hline
\multicolumn{2}{|c|}{ABB}&$-\sBRDM d{K_C}{K_B}{m_1}\la m_2m_3\ra$\\
\hline
\hline
$A$&$m_1$&$-\sBRDM d{K_C}{K_B}{q}\la m_2m_3\ra$\\
$/d$&$m_2$&$\sBRDM d{K_C}{K_B}{m_3}\la m_1q\ra$\\
&$m_3$&$\sBRDM d{K_C}{K_B}{m_2}\la qm_1\ra$\\
\hline
$B$&$m_1$&$0$\\
&$m_2$&$-\sBRDM d{K_C}{K_B}{m_1}\la qm_3\ra$\\
&$m_3$&$-\sBRDM d{K_C}{K_B}{m_1}\la m_2q\ra$\\
\hline
$C$&$m_1$&$-\sBRDM d{K_A}{K_B}{q}\la m_2m_3\ra$\\
&$m_2$&$\sBRDM d{K_A}{K_B}{q}\la m_3m_1\ra-\sBRDM d{K_C}{K_B}{m_1}\la qm_3\ra$\\
&$m_3$&$\sBRDM d{K_A}{K_B}{q}\la m_1m_2\ra-\sBRDM d{K_C}{K_B}{m_1}\la m_2q\ra$\\
\hline
\hline
\end{tabular}~~
%
%

\end{center}
\begin{center}\scriptsize
%
%
\begin{tabular}{|c|c|r|}
\hline
\hline
\multicolumn{2}{|c|}{ABd}&$\sBRDM d{K_C}{K_B}{m_2}\la dm_1\ra$\\
\hline
\hline
$A$&$m_1$&$\sBRDM d{K_C}{K_B}{m_2}\la dq\ra$\\
&$m_2$&$0$\\
&$d$&$\sBRDM d{K_C}{K_B}{m_2}\la qm_1\ra$\\
\hline
$B$&$m_1$&$-\sBRDM d{K_C}{K_B}{d}\la qm_2\ra$\\
&$m_2$&$\sBRDM d{K_C}{K_B}{q}\la dm_1\ra$\\
&$d$&$-\sBRDM d{K_C}{K_B}{m_1}\la m_2 q\ra$\\
\hline
$C$&$m_1$&$\sBRDM d{K_A}{K_B}{q}\la m_2d\ra-\sBRDM d{K_C}{K_B}{d}\la qm_2\ra$\\
&$m_2$&$-K_B^2\la dq\ra\la dm_1\ra$\\
&$d$&$\sBRDM d{K_A}{K_B}{q}\la m_1m_2\ra-\sBRDM d{K_C}{K_B}{m_1}\la m_2 q\ra$\\
\hline
\hline
\end{tabular}~~
%
%

\end{center}
\begin{center}\scriptsize
%
%
\begin{tabular}{|c|c|r|}
\hline
\hline
\multicolumn{2}{|c|}{ACd}&$K_B^2\la dm_1\ra\la dm_2\ra$\\
\hline
\hline
$A$&$m_1$&$K_B^2\la dm_2\ra\la dq\ra$\\
&$m_2$&$0$\\
&$d$&$K_B^2\la dm_2\ra\la qm_1\ra$\\
\hline
$B$&$m_1$&$\sBRDM d{K_A}{K_B}{q}\la m_2d\ra$\\
&$m_2$&$-\sBRDM d{K_C}{K_B}{q}\la dm_1\ra$\\
&$d$&$-\sBRDM d{K_A}{K_B}{m_2}\la qm_1\ra+\sBRDM d{K_C}{K_B}{m_1}\la m_2q\ra$\\
\hline
$C$&$m_1$&$0$\\
&$m_2$&$-K_B^2\la dm_1\ra\la m_2q\ra$\\
&$d$&$-K_B^2\la dm_1\ra\la qd\ra$\\
\hline
\hline
\end{tabular}~~
%
%
\begin{tabular}{|c|c|r|}
\hline
\hline
\multicolumn{2}{|c|}{BBd}&$\sBRDM d{K_A}{K_B}{d}\la m_1m_2\ra$\\
\hline
\hline
$A$&$m_1$&$\sBRDM d{K_C}{K_B}{m_2}\la dq\ra$\\
&$m_2$&$\sBRDM d{K_C}{K_B}{m_1}\la qd\ra$\\
&$d$&$-\sBRDM d{K_C}{K_B}{q}\la m_1m_2\ra$\\
\hline
$B$&$m_1$&$-\sBRDM d{K_C}{K_B}{d}\la qm_2\ra$\\
&$m_2$&$-\sBRDM d{K_C}{K_B}{d}\la m_1q\ra$\\
&$d$&$0$\\
\hline
$C$&$m_1$&$+\sBRDM d{K_A}{K_B}{m_2}\la dq\ra$\\
&$m_2$&$-\sBRDM d{K_A}{K_B}{m_1}\la qd\ra$\\
&$d$&$-\sBRDM d{K_A}{K_B}{q}\la m_1m_2\ra$\\
\hline
\hline
\end{tabular}\\ \ \\ \ \\
\end{center}
\begin{center}\scriptsize
%
%
\begin{tabular}{|c|c|r|}
\hline
\hline
\multicolumn{2}{|c|}{ABC}&
$\sBRDM d{K_A}{K_B}{m_3}\la m_1m_2\ra-\sBRDM d{K_C}{K_B}{m_1}\la m_2m_3\ra$\\
\hline
\hline
$A$&$m_1$&$\sBRDM d{K_A}{K_B}{m_3}\la qm_2\ra-\sBRDM d{K_C}{K_B}{q}\la m_2m_3\ra$\\
&$m_2$&$-K_B^2\la dm_3\ra\la m_1q\ra$\\
&$m_3$&$\sBRDM d{K_C}{K_B}{m_2}\la qm_1\ra$\\
\hline
$B$&$m_1$&$\sBRDM d{K_A}{K_B}{m_3}\la qm_2\ra$\\
&$m_2$&$\sBRDM d{K_A}{K_B}{m_3}\la m_1q\ra-\sBRDM d{K_C}{K_B}{m_1}\la qm_3\ra$\\
&$m_3$&$-\sBRDM d{K_C}{K_B}{m_1}\la m_2q\ra$\\
\hline
$C$&$m_1$&$-\sBRDM d{K_A}{K_B}{m_2}\la m_3q\ra$\\
&$m_2$&$K_B^2\la dm_1\ra\la qm_3\ra$\\
&$m_3$&$\sBRDM d{K_A}{K_B}{q}\la m_1m_2\ra-\sBRDM d{K_C}{K_B}{m_1}\la m_2q\ra$\\
\hline
$d$&$m_1$&$-\sBRDM d{K_A}{K_B}{m_2}\la m_3d\ra$\\
&$m_2$&$K_B^2\la dm_1\ra\la dm_3\ra$\\
&$m_3$&$\sBRDM d{K_C}{K_B}{m_2}\la dm_1\ra$\\
\hline
\hline
\end{tabular}
\normalsize
\end{center}
\smallskip
\begin{center}
Table 1: Numerators and Denominators For Conversion Factors $R^{xxx}_{qm}$
\end{center}
\bigskip 
The general effect of applying one of these conversion factors is to replace 
the ${\cal H}^4$ factor in the purely gluonic box-coefficient by ${\cal H}^3\tilde{\cal H}$,
where $\tilde{\cal H}$ is the factor appearing in the 'switched' purely gluonic box-coefficient, 
where leg $q$ is a negative helicity gluon and leg $m$ is a positive helicity gluon.
This is reminiscent of the behaviour of the MHV tree amplitudes, but in this case it
appears at the level of the box-coefficients.

So far we have only considered three-mass boxes.
As in the purely gluonic case, the two-mass and one-mass box-coefficients
for gluinos can be expressed as sums of three-mass box-coefficients. Given that
the factors appearing in the SWI are simply determined by the momenta of the
legs on which the supersymmetry generator acts, we see that, when expressed 
in terms of three-mass boxes, any SWI for (say) a two-mass easy box is just
a sum of the three-mass SWI and thus trivially satisfied. 
We have explicitly calculated the \npt  two-mass hard box-coefficients with 
two gluinos using quadruple cuts and verified the consistency of the two 
approaches. We have also used the 6-pt NMHV tree expressions of \cite{Bidder:2005in}
to calculate both singlet and non-singlet example 8-pt two-mass easy box-coefficients 
  using quadruple cuts. These results are also in agreement with 
those obtained by summing the appropriate three-mass coefficients.

\section{Beyond Two-Fermion Amplitudes} 

The conversion factors in table 1 can be compounded to generate amplitudes with arbitrary numbers of
external adjoint scalars and fermions. The first step is to note that the box-coefficient for
a diagram involving two external scalars can be obtained by simply squaring the conversion factor 
for the corresponding two gluino diagram,
$$
c^{xxx}(g^-,g^-,\phi^-_m,\phi^+_q,\ldots)=(R^{xxx}_{L(q)m})^2 c^{xxx}(g^-,g^-,g^-_m,g^+_q,\ldots).
\equn$$
For singlet two-gluino diagrams with only one possible route for the fermion, 
the corresponding two-scalar diagram  is obtained from the two gluino by replacing the single 
fermion line with a single scalar line. As we only have MHV and googly corners in the 
three-mass boxes, this simply gives us the square of the factor relating the two gluino
box-coefficient  to the gluonic. For two-gluino diagrams with two routes for the gluino, there are two-scalar
diagrams where the scalar takes one of these two routes and additionally there is 
a diagram with a fermionic loop. The first two diagrams give factors which are the
squares of the individual gluino factors, while explicit calculation shows that the last yields
precisely the cross-term that arises when the sum of the gluino terms is squared.
For the non-singlet diagrams,  explicit calculation again shows that the two scalar box-coefficients  are 
also simply obtained by squaring the relevant conversion factor.    
Recalling that we can express all of our box-coefficients 
in terms of the three-mass ones, we see that all the
two-scalar box-coefficients  are simply obtained by squaring the relevant factors in table 1.

To obtain SWI involving scalars we consider the action of a pair of supersymmetry generators
$Q_1$ and $Q_2$ that generate an $\NeqTwo$ subalgebra~\cite{SWI, Parke:1986jz}. 
The SWI then contain amplitudes involving
two flavours of gluino and a single flavour scalar. In $\NeqTwo$ terms it is natural to denote
scalar as $\phi^+\equiv\phi_{12}$ and $\phi^-\equiv\phi_{34}$. This notation is more compact than 
the full $\NeqFour$ flavour labelling, but care must be taken when counting the negative helicities 
required for a MHV vertex. In particular, replacing a $g^+$ by $\phi_{34}$ effectively introduces
an extra negative helicity. This is important in understanding the two-scalar ABC boxes, 
as all three of the diagrams shown below contribute to this two-scalar box-coefficient:

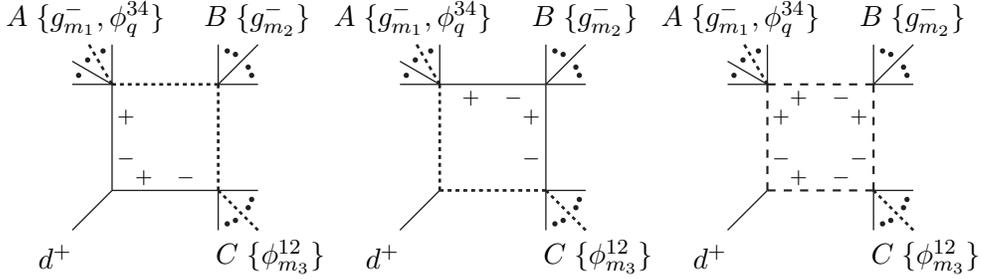
\begin{center}
\begin{picture}(120,100)(0,-5)
%
\SetWidth{1}
\DashLine(30,60)(70,60){1.5}
%
\DashLine(70,60)(70,20){1.5}
%
\SetWidth{0.5}
\Line(70,20)(30,20)
\Text(58,25)[c]{\scriptsize $-$}\Text(42,25)[c]{\scriptsize $+$}
%
\Line(30,20)(30,60)
\Text(32,32)[l]{\scriptsize $-$}\Text(32,48)[l]{\scriptsize $+$}
%
\Text(10,85)[c]{\hspace{20pt}\small $A \; \{g_{m_1}^- ,\phi^{34}_{q}\}$}
\Line(30,60)(15,60)
\Line(30,60)(15,68.7)
\SetWidth{1}
\DashLine(30,60)(21.3,75){1.5}
\SetWidth{0.5}
\Line(30,60)(30,75)
\Vertex(17.4,63.4){1}
\Vertex(20.8,69.2){1}
\Vertex(26.6,72.6){1}
%
\Text(90,85)[c]{\hspace{-10pt}\small $B \; \{g_{m_2}^-\}$}
\Line(70,60)(70,75)
\Line(70,60)(85,75)
\Line(70,60)(85,60)
\Vertex(73.4,72.6){1}
\Vertex(76.5,71.3){1}
\Vertex(81.3,66.5){1}
\Vertex(82.6,63.4){1}
%
\Text(90,-5)[c]{\small $C \; \{\phi^{12}_{m_3}\}$}
\Line(70,20)(85,20)
\SetWidth{1}
\DashLine(70,20)(85,5){1.5}
\SetWidth{0.5}
\Line(70,20)(70,5)
\Vertex(82.6,16.6){1}
\Vertex(81.3,13.5){1}
\Vertex(76.5,8.7){1}
\Vertex(73.4,7.4){1}
%
\Text(10,-5)[c]{\small $d^+ \; $}
\Line(30,20)(15,5)
\end{picture}
\begin{picture}(120,100)(0,-5)
%
\Line(30,60)(70,60)
\Text(42,55)[c]{\scriptsize $+$}\Text(58,55)[c]{\scriptsize $-$}
%
\Line(70,60)(70,20)
\Text(68,48)[r]{\scriptsize $+$}\Text(68,32)[r]{\scriptsize $-$}
%
\SetWidth{1}
\DashLine(70,20)(30,20){1.5}
%
\DashLine(30,20)(30,60){1.5}
\SetWidth{0.5}
%
\Text(10,85)[c]{\hspace{20pt}\small $A \; \{g_{m_1}^-, \phi^{34}_{q}\}$}
\Line(30,60)(15,60)
\Line(30,60)(15,68.7)
\SetWidth{1}
\DashLine(30,60)(21.3,75){1.5}
\SetWidth{0.5}
\Line(30,60)(30,75)
\Vertex(17.4,63.4){1}
\Vertex(20.8,69.2){1}
\Vertex(26.6,72.6){1}
%
\Text(90,85)[c]{\hspace{-10pt}\small $B \; \{g_{m_2}^-\}$}
\Line(70,60)(70,75)
\Line(70,60)(85,75)
\Line(70,60)(85,60)
\Vertex(73.4,72.6){1}
\Vertex(76.5,71.3){1}
\Vertex(81.3,66.5){1}
\Vertex(82.6,63.4){1}
%
\Text(90,-5)[c]{\small $C \; \{\phi^{12}_{m_3}\}$}
\Line(70,20)(85,20)
\SetWidth{1}
\DashLine(70,20)(85,5){1.5}
\SetWidth{0.5}
\Line(70,20)(70,5)
\Vertex(82.6,16.6){1}
\Vertex(81.3,13.5){1}
\Vertex(76.5,8.7){1}
\Vertex(73.4,7.4){1}
%
\Text(10,-5)[c]{\small $d^+ \; $}
\Line(30,20)(15,5)
\end{picture}
\begin{picture}(120,100)(0,-5)
%
\SetWidth{0.8}
\DashLine(30,60)(70,60)3
\Text(42,55)[c]{\scriptsize $+$}\Text(58,55)[c]{\scriptsize $-$}
%
\DashLine(70,60)(70,20)3
\Text(68,48)[r]{\scriptsize $+$}\Text(68,32)[r]{\scriptsize $-$}
%
\DashLine(70,20)(30,20)3
\Text(58,25)[c]{\scriptsize $-$}\Text(42,25)[c]{\scriptsize $+$}
%
\DashLine(30,20)(30,60)3
\SetWidth{0.5}
\Text(32,32)[l]{\scriptsize $-$}\Text(32,48)[l]{\scriptsize $+$}
%
\Text(10,85)[c]{\hspace{20pt}\small $A \; \{ g_{m_1}^- , \phi^{34}_q\}$}
\Line(30,60)(15,60)
\Line(30,60)(15,68.7)
\SetWidth{1}
\DashLine(30,60)(21.3,75){1.5}
\SetWidth{0.5}
\Line(30,60)(30,75)
\Vertex(17.4,63.4){1}
\Vertex(20.8,69.2){1}
\Vertex(26.6,72.6){1}
%
\Text(90,85)[c]{\hspace{-10pt}\small $B \; \{g_{m_2}^-\}$}
\Line(70,60)(70,75)
\Line(70,60)(85,75)
\Line(70,60)(85,60)
\Vertex(73.4,72.6){1}
\Vertex(76.5,71.3){1}
\Vertex(81.3,66.5){1}
\Vertex(82.6,63.4){1}
%
\Text(90,-5)[c]{\small $C \; \{ \phi^{12}_{m_3}\}$}
\Line(70,20)(85,20)
\SetWidth{1}
\DashLine(70,20)(85,5){1.5}
\SetWidth{0.5}
\Line(70,20)(70,5)
\Vertex(82.6,16.6){1}
\Vertex(81.3,13.5){1}
\Vertex(76.5,8.7){1}
\Vertex(73.4,7.4){1}
%
\Text(10,-5)[c]{\small $d^+ \; $}
\Line(30,20)(15,5)
\end{picture}
\end{center}
In these diagrams dotted lines represent scalars, while dashed lines represent fermions.

Next we consider the NMHV amplitudes with three non-gluonic  external legs. These box-coefficients 
are related to the purely gluonic ones by a pair of conversion factors:
$$
c^{xxx}(g^-,g^-,\phi^-_m,\gluinb^+_{q_1},\gluinb^+_{q_2}\ldots)
=R^{xxx}_{L(q_1)m}R^{xxx}_{L(q_2)m} c^{xxx}(g^-,g^-,g^-_m,g^+_{q_1},g^+_{q_2}\ldots),
\equn$$
$$
c^{xxx}(g^-,\gluino^-_{m_1},\gluino^-_{m_2},\phi^+_q,\ldots)
=R^{xxx}_{L(q)m_1} R^{xxx}_{L(q)m_2} c^{xxx}(g^-,g^-_{m_1},g^-_{m_2},g^+_q,\ldots).
\equn$$
For boxes with unique routings for the fermions, this result again follows directly from the form of the MHV amplitudes at 
each corner. For all boxes explicit calculation shows that the fermionic factors compound in this way.

Amplitudes involving four or more non-gluonic legs can now be generated directly from the appropriate SWI.
We define the 'level' of an amplitude to be the number of external fermions plus twice the number of 
external scalars.
Amplitudes with odd level will vanish and we use these as the starting points for our SWI. For example, 
acting with $Q_2$ on the level 3 amplitude, $A( \gluino_1^{1-}, g_2^{-},  g_3^- ,\gluinb_4^{1+},\gluinb_5^{2+},\ldots)$,
gives $A( \gluino_1^{1-},\gluino_2^{2-} ,  g_3^- ,\gluinb_4^{1+},\gluinb_5^{2+},\ldots)$ 
and $A( \gluino_1^{1-},g_2^- ,\gluino_3^{2-} ,\gluinb_4^{1+},\gluinb_5^{2+},\ldots)$ in terms of 
known amplitudes. We can now work systematically, level by level, to generate amplitudes with any
number of external scalars and fermions.

\section{Conclusions}

One-loop NMHV amplitudes in $\NeqFour$ gauge theory can be expressed in terms of MHV-deconstructible
diagrams and so can be evaluated using quadruple cuts and known MHV tree amplitudes. These amplitudes
also satisfy SWI which can be employed to minimize the number of diagrams that must be computed
explicitly. We have used these techniques to determine a set of conversion factors that relate
two-gluino box-coefficients  to purely gluonic ones. Analysis of quadruple cuts was then used to show how these
factors can be compounded to give two-scalar and scalar-gluino-gluino box-coefficients. Amplitudes involving more
external fermions/scalars then follow from SWI.  

\section{Acknowledgments}

We thank David Dunbar for more than useful
discussions.  This work was supported by a PPARC rolling grant.
SB would like to thank PPARC for a research studentship.


\begin{thebibliography}{99}

\bibitem{Witten:2003nn}
E.~Witten,
Commun.\ Math.\ Phys.\  {\bf 252}, 189 (2004),
[hep-th/0312171].

\bibitem{CSW}
F.~Cachazo, P.~Svrcek and E.~Witten,
JHEP {\bf 0409}, 006 (2004) [hep-th/0403047].

\bibitem{BDDK7}
Z.~Bern, V.~Del Duca, L.~J.~Dixon and D.~A.~Kosower,
Phys.\ Rev.\ D {\bf 71}, 045006 (2005),
[hep-th/0410224].

\bibitem{BDKn}
Z.~Bern, L.~J.~Dixon and D.~A.~Kosower,
hep-th/0412210.


\bibitem{BrittoUnitarity}
R.~Britto, F.~Cachazo and B.~Feng,
hep-th/0412103.


\bibitem{Cachazo:2004dr}
F.~Cachazo,
hep-th/0410077.


\bibitem{Britto:2004nj}
R.~Britto, F.~Cachazo and B.~Feng,
Phys.\ Rev.\ D {\bf 71}, 025012 (2005),
[hep-th/0410179].



\bibitem{Bidder:2004tx}
S.~J.~Bidder, N.~E.~J.~Bjerrum-Bohr, L.~J.~Dixon and D.~C.~Dunbar,
Phys.\ Lett.\ B {\bf 606}, 189 (2005),
[hep-th/0410296].


\bibitem{BBDP2005a}
S.\ J.\ Bidder, N.\ E.\ J.\ Bjerrum-Bohr,
D. C. Dunbar and W.\ B.\ Perkins
Phys.\ Lett.\ B {\bf 612}, 75 (2005),
[hep-th/0502028].

\bibitem{BrittoSQCD} 
R.~Britto, E.~Buchbinder, F.~Cachazo and B.~Feng,
hep-ph/0503132.

\bibitem{Bidder:2005in}
  S.~J.~Bidder, D.~C.~Dunbar and W.~B.~Perkins,
  \hepth{0505249}.

\bibitem{Huang:2005ve}
  Y.~t.~Huang,
  \hepth{0507117}.





\bibitem{SWI}
M.~T.~Grisaru, H.~N.~Pendleton and P.~van Nieuwenhuizen,
Phys.\ Rev.\ D {\bf 15}, 996 (1977);\\
M.~T.~Grisaru and H.~N.~Pendleton,
Nucl.\ Phys.\ B {\bf 124}, 81 (1977);\\
S.~J.~Parke and T.~R.~Taylor,
Phys.\ Lett.\ B {\bf 157}, 81 (1985),
[Erratum-ibid.\  B {\bf 174}, 465 (1986)].

\bibitem{TreeColour} 
F.A. Berends and W.T. Giele,
\npb{294}{700}{1987};\\ M.\ Mangano, S. Parke, and Z.\ Xu,
\npb{298}{653}{1988};\\ M.\ Mangano, \npb{309}{461}{1988}.
 
\bibitem{ManganoReview} M. Mangano and S.J. Parke, Phys.\ Rep.\
200:301 (1991).
                                               

\bibitem{ParkeTaylor}
S.J. Parke and T.R. Taylor,
Phys.\ Rev.\ Lett.\ 56:2459
(1986);\\
%
F.~A.~Berends and W.~T.~Giele,
Nucl.\ Phys.\ B {\bf 306}, 759 (1988).


                                                       
   

\bibitem{SpinorHelicity}
Z. Xu, D.-H.\ Zhang and L. Chang, \npb{291}{392}{1987}.



\bibitem{Colour}
Z. Bern and D.A.\ Kosower, \npb{362}{389}{1991}.


\bibitem{BDDKa}
Z. Bern, L.J. Dixon, D.C. Dunbar and D.A. Kosower,
\npb{425}{1994}{217}, \hepph{9403226}.

\bibitem{StringBased}
Z.~Bern and D.~A.~Kosower,
Phys.\ Rev.\ Lett.\  {\bf 66}, 1669 (1991).
Nucl.\ Phys.\ B {\bf 379}, 451 (1992);\\
Z.~Bern,
Phys.\ Lett.\ B {\bf 296}, 85 (1992);\\
Z.~Bern, D.~C.~Dunbar and T.~Shimada,
Phys.\ Lett.\ B {\bf 312}, 277 (1993),
[hep-th/9307001];\\
Z.~Bern and D.~C.~Dunbar,
Nucl.\ Phys.\ B {\bf 379}, 562 (1992);\\
D.C. Dunbar and P.S. Norridge, Class. Quantum Grav. {\bf 14}, 351 {(1997)},
[hep-th/9512084].


\bibitem{BDDKb}
Z. Bern, L.J. Dixon, D.C. Dunbar and D.A. Kosower,
\npb{435}{1995}{59}, \hepph{9409265}.


\bibitem{MultiLoop}
 Z. Bern, J.S. Rozowsky and B. Yan, Phys.\ Lett.\ B401:273 (1997),
[hep-ph/9702424];
\\
 C.~Anastasiou, Z.~Bern, L.~J.~Dixon and D.~A.~Kosower,
Phys.\ Rev.\ Lett.\  {\bf 91}, 251602 (2003),
[hep-th/0309040];
\\Z.~Bern, L.~J.~Dixon and V.~A.~Smirnov,
  hep-th/0505205.



\bibitem{Siegel}
W.~Siegel,
Phys.\ Lett.\ B {\bf 84}, 197 (1979);\\
D.~M.~Capper, D.~R.~T.~Jones and P.~van Nieuwenhuizen,
Nucl.\ Phys.\ B {\bf 167}, 479 (1980);\\
L.~V.~Avdeev and A.~A.~Vladimirov,
Nucl.\ Phys.\ B {\bf 219}, 262 (1983).

\bibitem{KST}
Z.~Kunszt, A.~Signer and Z.~Trocsanyi,
Nucl.\ Phys.\ B {\bf 411}, 397 (1994),
[hep-ph/9305239].




\bibitem{CSW:matter}
G.~Georgiou and V.~V.~Khoze,
JHEP {\bf 0405}, 070 (2004),   \,\,\,\,\,\,\,\,\,\,\,\,\,\,\,\,\,\,\,
\hbox{[hep-th/0404072];}\\
%
%
J.~B.~Wu and C.~J.~Zhu,
JHEP {\bf 0409}, 063 (2004) [hep-th/0406146];\\
%
%
X.~Su and J.~B.~Wu,
hep-th/0409228.
%

\bibitem{CSW:massive}
L.~J.~Dixon, E.~W.~N.~Glover and V.~V.~Khoze,
JHEP {\bf 0412}, 015 (2004),
[hep-th/0411092];\\
Z.~Bern, D.~Forde, D.~A.~Kosower and P.~Mastrolia,
[hep-ph/0412167];\\
S.~D.~Badger, E.~W.~N.~Glover and V.~V.~Khoze,
JHEP {\bf 0503}, 023 (2005),
[hep-th/0412275].


\bibitem{Britto:2004ap}
R.~Britto, F.~Cachazo and B.~Feng,
Nucl.\ Phys.\ B {\bf 715}, 499 (2005),
[hep-th/0412308]; \\
R.~Britto, F.~Cachazo, B.~Feng and E.~Witten,
hep-th/0501052.

\bibitem{Britto:2005dg}
R.~Britto, B.~Feng, R.~Roiban, M.~Spradlin and A.~Volovich,
Phys.\ Rev.\ D {\bf 71}, 105017 (2005),
[hep-th/0503198].



\bibitem{Luo:2005rx}
M.~Luo and C.~Wen,
JHEP {\bf 0503}, 004 (2005),
[hep-th/0501121].

\bibitem{Bedford:2005yy}
J.~Bedford, A.~Brandhuber, B.~Spence and G.~Travaglini,
\,\,\,\,\,\,\,\,\,\,\,\,\,\,\,\,\,\,\,\,
 \hbox{hep-th/0502146.}


\bibitem{Cachazo:2005ca}
 F.~Cachazo and P.~Svrcek,
  hep-th/0502160.


\bibitem{Bjerrum-Bohr:2005xx}
  N.~E.~J.~Bjerrum-Bohr, D.~C.~Dunbar and H.~Ita,
  hep-th/0503102.


\bibitem{EllisSexton}
R.~K.~Ellis and J.~C.~Sexton,
  Nucl.\ Phys.\ B {\bf 269}, 445 (1986);\\
M.B.\ Green, J.H.\ Schwarz and L.\ Brink,
 Nucl.\ Phys.\ B198:472 (1982).

\bibitem{FiveGluon}Z. Bern, L. Dixon and D.A.\ Kosower, Phys.\ Rev.\ Lett.\
70:2677 (1993).


\bibitem{Bern:1994fz}
  Z.~Bern, L.~J.~Dixon and D.~A.~Kosower,
  Nucl.\ Phys.\ B {\bf 437}, 259 (1995),
  [hep-ph/9409393].


\bibitem{Kunszt:1994tq}
  Z.~Kunszt, A.~Signer and Z.~Trocsanyi,
  Phys.\ Lett.\ B {\bf 336}, 529 (1994),
  [hep-ph/9405386].


\bibitem{Glover:1996eh}
  E.~W.~N.~Glover and D.~J.~Miller,
  Phys.\ Lett.\ B {\bf 396}, 257 (1997),
  [hep-ph/9609474].

\cite{Bern:2005cq}
\bibitem{Bern:2005cq}
  Z.~Bern, L.~J.~Dixon and D.~A.~Kosower,
  hep-ph/0507005.




\bibitem{Bern:2005hh}
  Z.~Bern, N.~E.~J.~Bjerrum-Bohr, D.~C.~Dunbar and H.~Ita,
  hep-ph/0507019.


\bibitem{Brandhuber:2004yw}
A.~Brandhuber, B.~Spence and G.~Travaglini,
 Nucl.\ Phys.\ B {\bf 706}, 150 (2005), [hep-th/0407214].


\bibitem{Quigley:2004pw}
C.~Quigley and M.~Rozali,
 JHEP {\bf 0501}, 053 (2005), [hep-th/0410278].

\bibitem{Bedford:2004py}
J.~Bedford, A.~Brandhuber, B.~Spence and G.~Travaglini,
 Nucl.\ Phys.\ B {\bf 706}, 100 (2005),
 [hep-th/0410280].

\bibitem{Eden}
R.J. Eden, P.V. Landshoff, D.I. Olive, J.C. Polkinghorne, {\it
The Analytic S Matrix}, (Cambridge University Press, 1966).
 

\bibitem{GeneralizedCuts}
Z.~Bern, L.~J.~Dixon and D.~A.~Kosower,
Nucl.\ Phys.\ B {\bf 513}, 3 (1998),
[hep-ph/9708239];
JHEP {\bf 0001}, 027 (2000),
[hep-ph/0001001];\\
Z.~Bern, L.~J.~Dixon and D.~A.~Kosower,
JHEP {\bf 0408}, 012 (2004),
[hep-ph/0404293]; \\
Z. Bern, L. Dixon, D.C. Dunbar and D.A. Kosower,
\plb{394}{1997}{105}, \hepth{9611127}.


\bibitem{BBDP}
S.~J.~Bidder, N.~E.~J.~Bjerrum-Bohr, D.~C.~Dunbar and
W.~B.~Perkins,
Phys.\ Lett.\ B {\bf 608}, 151 (2005),
[hep-th/0412023].




                                                                               



\bibitem{Bern:2005hs}
  Z.~Bern, L.~J.~Dixon and D.~A.~Kosower,
  hep-th/0501240.


\bibitem{Bern:2005ji}
  Z.~Bern, L.~J.~Dixon and D.~A.~Kosower,
  hep-ph/0505055.

 \bibitem{Nair}
V.~P.~Nair,
Phys.\ Lett.\ B {\bf 214}, 215 (1988).
 



\bibitem{Parke:1986jz}
 S.~J.~Parke,
FERMILAB-CONF-86-044-T
{\it Invited talk given at Int. Symp. on Particle and Nuclear Physics, Beijing, China, Sep 2-7, 1985}. 







\end{thebibliography}
\end{document}